\documentclass[sigconf,nonacm]{acmart}

\acmConference[ICSE-NIER '26]{New Ideas and Emerging Results}{Apr 2026}{Rio de Janeiro, Brazil}

\acmBooktitle{Proceedings of the 48th IEEE/ACM International Conference on Software Engineering: NIER (ICSE-NIER '26)}

\usepackage{booktabs}   
\usepackage{subcaption}
\usepackage{microtype}
\usepackage[utf8]{inputenc}
\usepackage{algorithm}
\usepackage{algorithmic}
\usepackage{enumitem}
\title{Model Discovery and Graph Simulation: A Lightweight Gateway to Chaos Engineering}

\author{Anatoly A. Krasnovsky}
\orcid{0000-0001-6842-7340}
\affiliation{
  \department{Department of Computer Science and Engineering}
  \institution{Innopolis University}
  \city{Innopolis}
  \country{Russia}
}
\affiliation{
  \institution{MB3R Lab}
  \city{Innopolis}
  \postcode{420500}
  \country{Russia}}

\renewcommand{\shortauthors}{Krasnovsky}

\begin{document}

%
\begin{abstract}
Chaos engineering reveals resilience risks but is expensive and operationally risky to run broadly and often. Model-based analyses can estimate dependability, yet in practice they are tricky to build and keep current because models are typically handcrafted. We claim that a simple connectivity-only \emph{topological} model—just the service-dependency graph plus replica counts—can provide fast, low-risk availability estimates under fail-stop faults. To make this claim practical without hand-built models, we introduce \emph{model discovery}: an automated step that can run in CI/CD \emph{or} as an observability-platform capability, synthesizing an explicit, analyzable model from artifacts teams already have (e.g., distributed traces, service-mesh telemetry, configs/manifests)—\emph{providing an accessible gateway for teams to begin resilience testing}.
As a proof by instance on the DeathStarBench Social Network, we extract the dependency graph from Jaeger and estimate availability across two deployment modes and five failure rates. The discovered model closely tracks live fault-injection results; with replication, median error at mid-range failure rates is near zero, while no-replication shows signed biases consistent with excluded mechanisms. These results create two opportunities: first, \emph{to triage and reduce the scope of expensive chaos experiments in advance}, and second, \emph{to generate real-time signals on the system's resilience posture as its topology evolves}, preserving live validation for the most critical or ambiguous scenarios.
\end{abstract}

\keywords{microservices, resilience, model discovery, graph simulation, chaos engineering}

\maketitle

\section{Introduction}
Microservices accelerate delivery and scale, but implicit inter-service dependencies can turn local faults into cascading failures. Resilience practice combines patterns such as circuit breakers, retries, and replication with live fault injection; these techniques surface failure modes but broad experimentation remains time-consuming, operationally risky, and hard to run continuously \cite{fowler_circuitbreaker,grottke_fighting_bugs_remove,basiri_chao_engineer,heorhiadi_gremlin_systemati}. 

A complementary path is to reason from models. Classical dependability and stochastic analysis provide tools to estimate availability and guide design decisions without disrupting production \cite{avizienis_basic_concept,kleijnen_design_and_analysis,rausand_system_reliabili}. For microservices, model-based analyses (e.g., PRISM encodings of resiliency patterns) quantify how retries and circuit breakers affect success probability under component failures \cite{kwiatkowska_prism40_verificat,mendonc_model}. In practice, however, such models are typically handcrafted, which does not scale with fast-moving stacks. Empirical studies report that maintaining architecture and analysis models in microservice settings is challenging and often remains largely manual~\cite{kleehaus2019challenges}. Practitioners further describe the code-to-architecture mapping needed to keep such models consistent as high‑effort and largely manual~\cite{ali2018architecture}, and even in classical dependability/performance contexts hand‑built Markov models are recognized as tedious and error‑prone~\cite{brameret2015partialmc}.

\textbf{Claim.} A simple connectivity-only \emph{topological} model—blocking call edges plus replica counts—can provide fast, low-risk estimates of endpoint availability under a fail-stop assumption, providing a fast way to triage where expensive chaos adds the most value.

\textbf{Novelty.} Our contribution is not a new modeling formalism but the \emph{automation} of model synthesis from passive artifacts, making an analyzable model continuously available without hand-crafting. We estimate endpoint availability as the probability that a request completes without transport errors, timeouts, or server-side failures.

\textbf{Proposal: model discovery.} To make the claim practical without hand-built models, we introduce \emph{model discovery}: an automated, source-agnostic step that can run in CI/CD \emph{or} as an observability-platform capability. It synthesizes an explicit, analyzable service-dependency model from artifacts teams already have: distributed tracing (which captures causal call edges \cite{sigelman2010dapper,opentelemetry-overview}), service-mesh telemetry/config, configs/manifests (e.g., Terraform configurations and Kubernetes manifests that declare components and resource dependencies \cite{terraform-graph,k8s-service,soldani2023offline}), and API contracts. For traces, common platforms already expose a dependency graph that can be exported and analyzed \cite{jaeger-docs-faq,jaeger-intro}. 

\textbf{Proof by instance.} We validate a traces-only instantiation on the DeathStarBench Social Network benchmark \cite{ganopenbench}, extracting the dependency graph from Jaeger and evaluating across replication regimes and failure rates. As we show later, predictions from this simple graph-based model closely track live fault-injection outcomes. 

\textbf{Contributions.}
\begin{itemize}[nosep]
  \item Formulate \emph{model discovery} as a source-agnostic pipeline step (or a platform module) that builds analyzable models (graphs; optionally extended abstractions) from existing artifacts.
  \item Realize a \emph{traces-only} path that discovers a dependency graph and estimates endpoint availability under fail-stop faults.
  \item Empirically show close agreement with live chaos on DeathStarBench across failure rates and replication regimes, indicating that connectivity+replication captures much of the resilience signal.
\end{itemize}

\section{Approach}

\subsection{Model}
We model a microservice application as a typed, directed dependency graph
$G=(V,E)$ where each $v\in V$ is a service and each edge $u\!\to\!v\in E$ denotes a \emph{blocking} call required for request completion.
Each service $v$ carries a replica count $r(v)\!\in\!\mathbb{N}_{\ge 1}$; user-visible \emph{entrypoints} (endpoints) are tagged in $V$.
Under a fail-stop fault model and synchronous calls, a request to an entrypoint succeeds iff every required service on at least one feasible call chain is \emph{alive}, and each such service has at least one live replica. This makes $G$ (plus $r(\cdot)$ and entrypoints) an informative predictor of availability.

\subsection{Discovery pipeline}
\textbf{Goal.} Synthesize $G$ automatically from artifacts teams already have—either as a CI/CD step or an observability‑platform capability.

\noindent\textbf{Inputs (any one is sufficient; fusion is optional):}
\begin{itemize}
  \item \textbf{Distributed tracing} (e.g., OpenTelemetry/Jaeger): traces expose causal call edges suitable for service‑dependency graphs. \cite{sigelman2010dapper,opentelemetry-overview,jaeger-docs-faq}
  \item \textbf{Service‑mesh telemetry/config}: L7 routing and policies (one‑hop edges) align with Jaeger’s dependency graph semantics.
  \item \textbf{Configs/manifests}: Infrastructure as Code (IaC). Terraform computes resource graphs (\texttt{terraform graph}); Kubernetes Services/selectors bind traffic to backends—both provide structural dependencies that can seed $G$. \cite{terraform-graph,k8s-service}
  \item \textbf{API contracts/IDLs} (OpenAPI/Thrift/Proto): precise interface requirements (optionally linked to clients).
  \item \textbf{SLO-as-Code (SLOaC)}: Declarative SLO files (YAML) that name critical
endpoints/user journeys and their target availability/error budgets. We treat SLOs
as first-class metadata on entrypoints and use them to prioritize analysis (and later
to compare $\hat R_{\text{model}}$ against targets).

\end{itemize}

\noindent\textbf{Pipeline:}
\begin{enumerate}
  \item \emph{Ingest}: parse the chosen source(s) (traces, mesh, IaC, API specs).
  \item \emph{Normalize}: canonicalize service identities; collapse replicas to one node with $r(v)$; harmonize protocol tags.
  \item \emph{Edge extraction}: emit $u\!\to\!v$ when evidence indicates a blocking dependency (trace causality; mesh flow; IaC binding; client stubs).
  \item \emph{Typing \& pruning}: tag entrypoints; if SLOaC is present, attach per-endpoint targets and order/prioritize evaluation accordingly; mark optional edges when detectable; prune obvious non-blocking paths for evaluated endpoints.
  \item \emph{Provenance \& checks}: record per‑edge source(s); run sanity checks (dangling nodes, degree anomalies); output a short report and the final graph model $G$ used for estimation.
\end{enumerate}

\paragraph*{Note on IaC as a source.}
Beyond vendor docs, prior work shows full architectures mined \emph{offline} from Kubernetes deployments/Mani\-fests, reinforcing IaC/manifests as a first‑class discovery input. \cite{soldani2023offline}

\paragraph*{Note on trace-derived service graphs.} Trace-derived service graphs (e.g., Jaeger Dependencies) are service-level and may blur optional vs.\ blocking edges or conflate per-endpoint paths; we therefore record per-edge provenance and allow optional-edge heuristics (e.g., frequency thresholds) and endpoint-scoped pruning when such signals are available.

\subsection{Availability estimation}

\begin{algorithm}[t]
\caption{Availability estimation from $G$}
\label{alg:availability}
\begin{algorithmic}[1]
\REQUIRE graph $G=(V,E)$, replica map $r(\cdot)$, failure fraction $p_\mathrm{fail}$, number of samples $N$ \textit{(RNG seed $\sigma$ optional)}, entrypoints $\mathcal{E}$
\ENSURE $\hat R_\text{model}(e,p_\mathrm{fail})$ for all $e\in\mathcal{E}$
\STATE Initialize RNG with $\sigma$ if provided
\STATE \textbf{Precompute:} contract strongly connected components (SCCs) of $G$ to obtain a DAG $\widehat{G}$; remove nodes with $r(v)=0$ \textit{(self-loops vanish under contraction)}
\FOR{each $i \in \{1,\dots,N\}$}
    \STATE \COMMENT{sample failures}
    \STATE Independently fail each replica with probability $p_\mathrm{fail}$
    \STATE A service $v$ is alive iff any of its $r(v)$ replicas survives
    \STATE Let $V_i^\mathrm{alive}$ be the set of alive services
    \STATE \COMMENT{reachability on $\widehat{G}$}
    \FOR{each $e \in \mathcal{E}$}
        \STATE Mark success if there exists a \emph{feasible call chain} from $e$ in the subgraph of $\widehat{G}$ induced by $V_i^\mathrm{alive}$, where a feasible call chain is any directed path of \emph{blocking} edges from $e$ to a leaf
        \STATE \COMMENT{Disconnected components not containing $e$ are ignored implicitly}
    \ENDFOR
\ENDFOR
\STATE \COMMENT{Per-sample reachability is $\mathcal{O}(|\widehat{V}|+|\widehat{E}|)$ via one BFS/DFS on the alive subgraph}
\STATE Return per-$e$ mean (and SD) over $N$ samples
\end{algorithmic}
\end{algorithm}

Given failure fraction $p_{\text{fail}}$ and a chosen deployment mode (replica map $r(\cdot)$), we estimate endpoint availability using the procedure in Algorithm~\ref{alg:availability}. For each endpoint $e$, feasibility is restricted to call paths actually observed in successful workload traces; edges not on such paths are ignored. We do not further classify optional vs. required beyond this.

\begin{enumerate}
  \item Sample a failure set $F\subseteq V$ by independently removing a $p_{\mathit{fail}}$ fraction of service \emph{instances}; a service $v$ remains alive if at least one of its $r(v)$ replicas is alive.
  \item For each entrypoint $e$, declare success if there exists a path from $e$ to all required services in $G\setminus F$ under the blocking‑edge semantics (reachability in the alive subgraph).
  \item Repeat for a small number of seeds and report $\hat{R}_{\text{model}}(e,p_{\mathit{fail}})$ as the mean success rate (with SD).
\end{enumerate}
This procedure avoids hand‑built analytic models while remaining fast enough for CI and cheap enough for continuous observability.

\paragraph*{Why simulation?} Exact network-reliability computation is intractable in general graphs; lightweight Monte-Carlo gives CI-speed estimates suitable for CI/CD \cite{paredes2019principled}.

\paragraph*{Why connectivity can suffice?}
Under fail‑stop faults and synchronous calls, end‑to‑end success reduces to (i) reachability in the alive subgraph and (ii) replication at required services. Thus a simple graph plus replica counts often captures most of the resilience signal; extensions (gray failures, correlated shocks, async/event flows) are orthogonal and can be layered later.

\section{System and Experimental Setup}
We evaluate on the DeathStarBench Social Network \cite{ganopenbench}. A constant\-‑throughput workload is generated with \texttt{wrk2} \cite{wrk2} against the three user‑facing endpoints using the canonical 1:3:6 mix \cite{latenseer-socc23}. The live success rate is
\[
R_{\text{live}} = 1 - \frac{\#\text{HTTP 5xx} + \#\text{socket errors} + \#\text{timeouts}}{\#\text{total requests}} \, ,
\]
i.e., we count transport‑level errors and timeouts in addition to server‑side failures. We assume fail‑stop faults at the service‑contai\-ner granularity and consider two deployment modes (\emph{norepl}, \emph{repl}) across failure fractions \(p_{\mathit{fail}}\in\{0.1,0.3,0.5,0.7,0.9\}\), a range chosen to capture the system's behavior from the onset of significant degradation—defined here as the first step beyond the single-fault limit of the norepl baseline—through to catastrophic failure.

We drive load with a fixed rate and short windows:

\verb|wrk2 -t 2 -c 64 -d 30s -R 300 -L -s|; where
\verb|-t| threads, \verb|-c| connections, \verb|-d| duration, and \verb|-R| rate.

\emph{Why not the naive ``non-200'' count?}
HTTP semantics treat many non-200 statuses as \emph{not} service unavailability
(e.g., 3xx redirection; many 4xx client-side issues), while true outages often
surface as \emph{no status} at all (socket errors, timeouts). Following SRE
practice, we therefore count \textbf{server-side} errors (5xx) and \textbf{transport-level}
failures (socket errors, timeouts) as ``bad'' events, and everything else as ``good''. This avoids inflating availability by silently dropping attempts that never produced a status code and avoids penalizing expected 3xx or user-caused 4xx.

Execution is automated as a GitHub Actions matrix yielding \emph{250 jobs}. Each job runs the full pipeline for its configuration and produces one aggregated data point used in our analysis (Figures~1–2). \emph{Within a job}, the estimator executes \textbf{4{,}500{,}000} graph simulations (five failure fractions, \texttt{--samples} \(=\) \textbf{900\,000} each) and the chaos harness performs \textbf{450} (limited to  \texttt{timeout-minutes:360} per job) live measurement windows; summaries are emitted per job. To demonstrate single‑source viability, discovery uses \emph{traces only}: we export Jaeger’s \emph{Dependencies} summary and convert it into the typed, directed service graph \cite{sigelman2010dapper,opentelemetry-overview,jaeger-docs-faq,jaeger-intro}. The artifact and CI scripts are publicly archived \anon{\cite{github_krasnovsky}}.

\section{Results}

\begin{figure}[t]
  \centering
  \includegraphics[width=\linewidth]{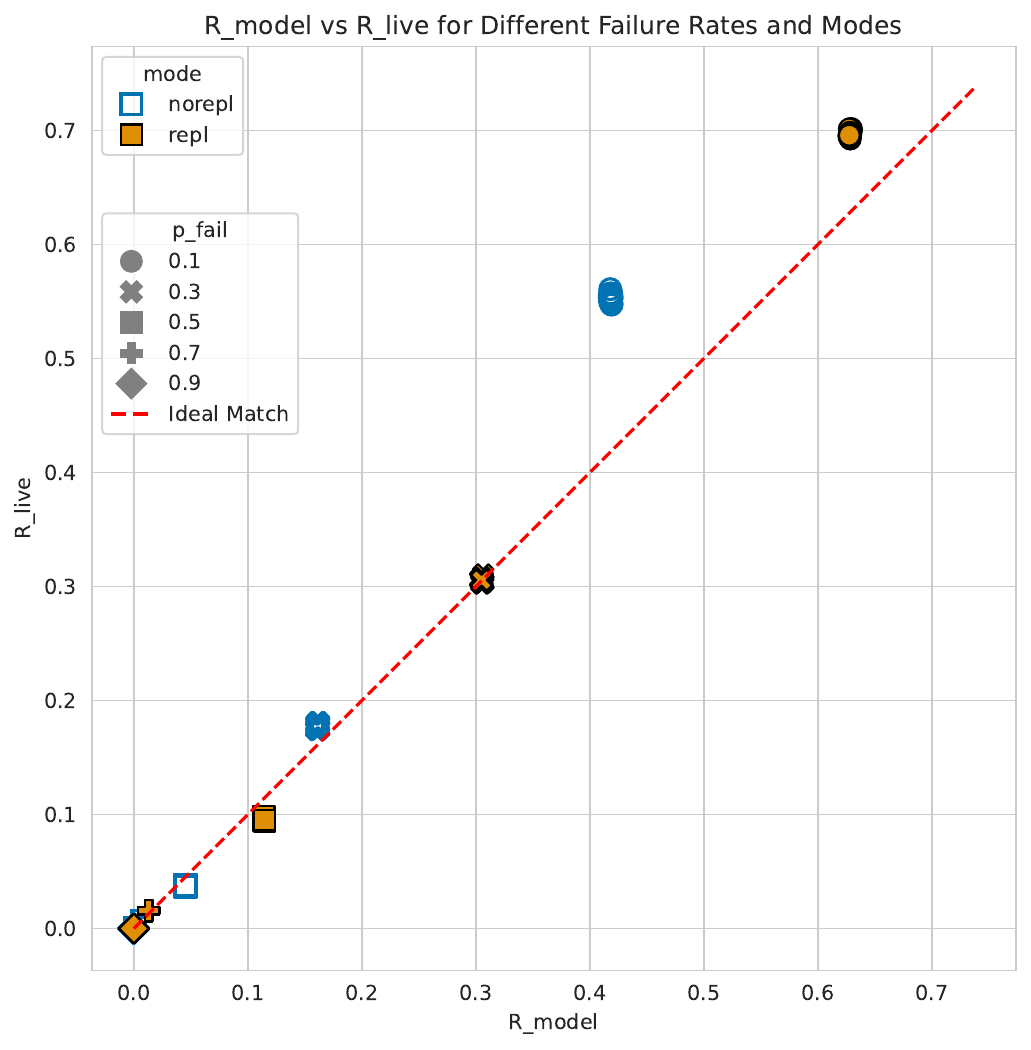}
  \Description{Scatter of $\hat R_\text{model}$ versus $R_\text{live}$ across
  two deployment modes (norepl, repl) and five failure fractions $p_\mathrm{fail}\in\{0.1,0.3,0.5,0.7,0.9\}$.
  Replicated points lie higher and closer to the dashed $y{=}x$ diagonal than no-replication.}
  \caption{$R_{\text{model}}$ vs.\ $R_{\text{live}}$ across all conditions; dashed $y{=}x$ shows ideal agreement. \textbf{Replication effect captured:} replicated points lie higher and closer to the diagonal than no‑replication. Points are CI‑job aggregates (4.5M simulations $+$ 450 live windows per job).}
  \label{fig:model-vs-live}
\end{figure}
\begin{figure}[t]
  \centering
  \includegraphics[width=\columnwidth]{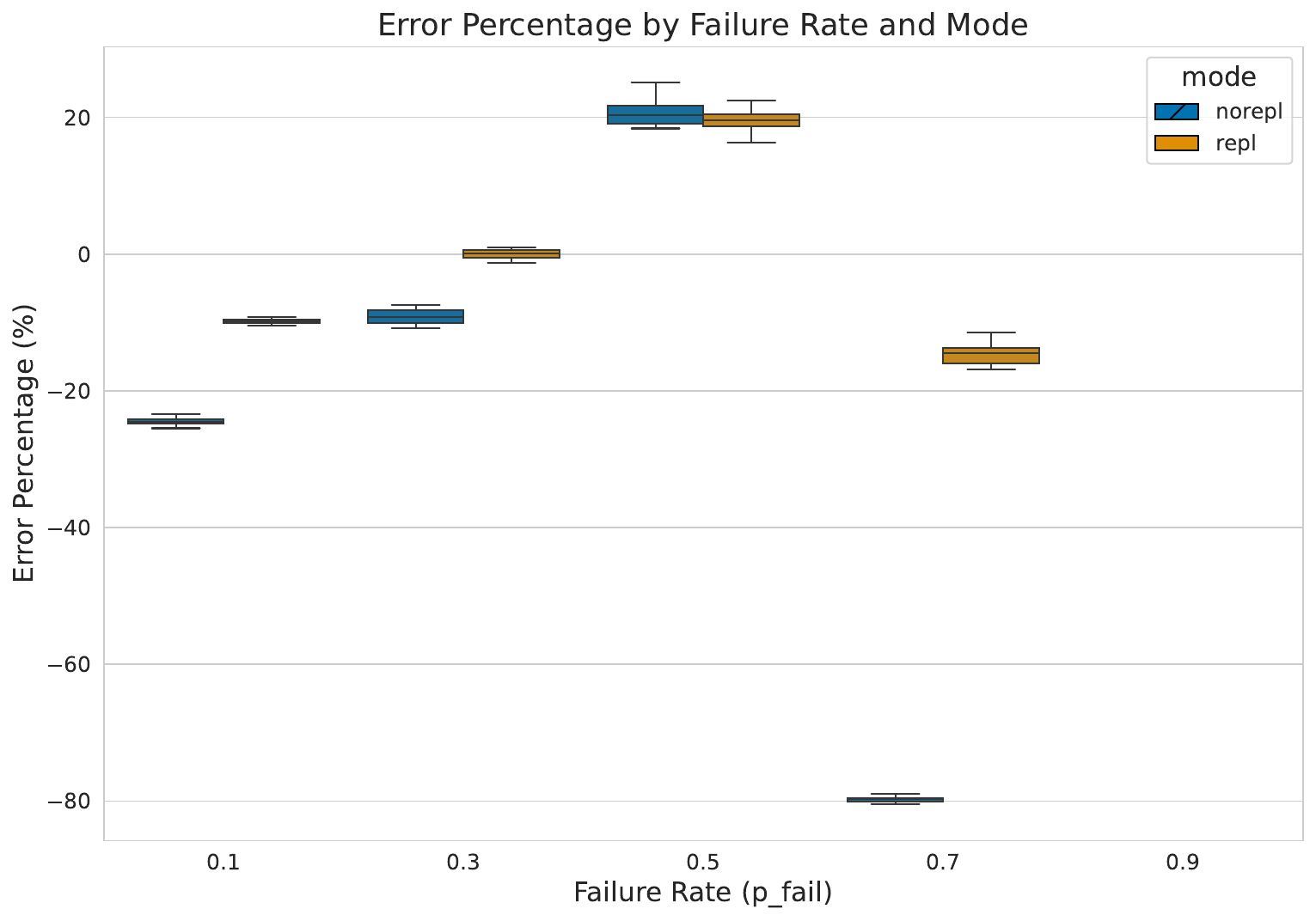}
  \Description{Line/marker chart of relative error $\Delta\% = 100\times|R_\text{model}-R_\text{live}|/R_\text{live}$
  across failure fractions for norepl and repl. Norepl shows larger deviations at low and mid $p_\mathrm{fail}$,
  while repl remains small and near-zero at mid-range.}
  \caption{Error percentage ($\Delta\%$) by failure rate and deployment mode. 
  Summarizes the relative model--live discrepancy per condition using the same 
  CI-job aggregates as Fig.~1 and Table~1.}
  \label{fig:error-percentage}
\end{figure}

The discovered topological model tracks live outcomes closely and faithfully reflects the deployment regime. The model correctly captures the significant resilience benefit of replication, which, at a mid-range failure rate ($p_\mathrm{fail}=0.3$), drives the prediction error to near zero. In contrast, scenarios without replication exhibit systematic signed biases (e.g., -24.4\% at $p_\mathrm{fail}=0.1$, +20.7\% at $p_\mathrm{fail}=0.5$). As the analysis below confirms, this near-zero error is statistically indistinguishable from reality, while the other biases are statistically significant. Figure~\ref{fig:model-vs-live} visualizes the overall agreement; Table~\ref{tab:agg} and Figure~\ref{fig:error-percentage} detail the results per condition.

A rigorous statistical analysis supports these observations (overall Pearson $r \approx 0.992$). In the key scenario—with replication enabled at $p_\mathrm{fail}=0.3$—a two-sample t-test confirms the difference between model and live results is \emph{statistically insignificant} ($p \approx 0.973$). In all other conditions, however, the observed biases are \emph{statistically significant} after applying a Benjamini-Hochberg correction for multiple comparisons. These biases are not random error but are consistent with the model's intentional simplifications, pointing to the measurable impact of excluded phenomena like retry mechanisms at low failure rates and cascading timeouts at high failure rates.

\begin{table}[t]
  \centering
  \small
  \caption{Aggregate resilience by mode and failure fraction (mean $\pm$ SD).}
  \label{tab:agg}
  \begin{tabular}{lccccc}
    \hline
    Mode & $p_{\mathit{fail}}$ & $R_{\text{model}}$ & $R_{\text{live}}$ \\
    \hline
    norepl & 0.1 & 0.4182 $\pm$ 0.0005 & 0.5533 $\pm$ 0.0031 \\
    norepl & 0.3 & 0.1613 $\pm$ 0.0005 & 0.1775 $\pm$ 0.0021 \\
    norepl & 0.5 & 0.0454 $\pm$ 0.0002 & 0.0376 $\pm$ 0.0006 \\
    norepl & 0.7 & 0.0014 $\pm$ 0.0000 & 0.0067 $\pm$ 0.0000 \\
    norepl & 0.9 & 0.0000 $\pm$ 0.0000 & 0.0000 $\pm$ 0.0000 \\
    repl   & 0.1 & 0.6281 $\pm$ 0.0005 & 0.6969 $\pm$ 0.0026 \\
    repl   & 0.3 & 0.3054 $\pm$ 0.0007 & 0.3054 $\pm$ 0.0017 \\
    repl   & 0.5 & 0.1145 $\pm$ 0.0004 & 0.0958 $\pm$ 0.0011 \\
    repl   & 0.7 & 0.0132 $\pm$ 0.0001 & 0.0155 $\pm$ 0.0003 \\
    repl   & 0.9 & 0.0000 $\pm$ 0.0000 & 0.0000 $\pm$ 0.0000 \\
    \hline
  \end{tabular}
\end{table}

\section{Discussion and Related Work}
Our results show that a connectivity‑only model augmented with replica counts captures most of the resilience signal and \emph{faithfully follows the deployment regime}: switching from \textit{norepl} to \textit{repl} raises both $R_{\text{live}}$ and $R_{\text{model}}$ and drives mid‑range error to $\approx 0$. Residual, mode‑dependent biases are small and interpretable: they are consistent with phenomena the model intentionally leaves out (e.g., partial/gray failures, optional or non‑blocking edges, request‑mix drift, retries/load‑shedding, and queueing effects), which can cause live behavior to be slightly better at low $p_{\mathit{fail}}$ and slightly worse around $0.5$ \cite{huang_grey_failure}. The success metric we use (HTTP~5xx \emph{plus} socket errors and timeouts) also tightens the live denominator relative to “non‑200” counts, explaining part of the underestimation in \textit{norepl}; nevertheless the pattern is stable across seeds and collapses to zero at the extreme, matching the model (Fig.~\ref{fig:model-vs-live}, Table~\ref{tab:agg}). 

This work complements two lines of prior art. First, chaos engineering and systematic fault‑injection frameworks surface real failure modes but are costly to run broadly \cite{basiri_chao_engineer,heorhiadi_gremlin_systemati,nygard_release_it,microsoft_resilienc_pattern}; our discovered model triages where such experiments add the most value. Second, model‑based analyses provide principled estimates but typically rely on handcrafted abstractions \cite{avizienis_basic_concept,rausand_system_reliabili,kleijnen_design_and_analysis,kwiatkowska_prism40_verificat,mendonc_model}. Our contribution is to \emph{automate} model construction from artifacts teams already have—traces and observability in particular—leveraging decades of tracing practice \cite{sigelman2010dapper,li_enjoy_observe}. This positions model discovery alongside recent work that mines architecture/topology from logs or manifests \cite{soldani_explain_microserv,soldani2023offline} and complements performance/resilience tooling (e.g., microservice simulation and automated robustness profiling) \cite{zhang_mqsim,giamattei_automated_functiona,yang_microres_versatile_resilience}.

\paragraph*{Positioning.} Microservice simulators (e.g., µqSim) target performance under complex queuing and resource dynamics; MicroRes profiles live systems via targeted degradation to surface robustness issues. Our goal differs: we show how far a \emph{discovered, connectivity-only} model can go for availability estimation without live injection. Mining-based works \cite{soldani2023offline} recover architecture/topology from logs/manifests; we \emph{use} such recovered graphs (from traces/mesh/IaC) to produce quantitative resilience estimates and continuous posture signals.

\section{Future Plans} \label{future}
Building on the observed agreement between the discovered model and live results—and the fact that the model faithfully reflects the replication regime—we outline the concrete steps to turn this work into a full paper. \emph{Scope remains the same: a lightweight, automatically discovered topological model that teams can run in CI/CD or as an observability capability.}

\noindent\textbf{Milestones and success criteria.}
\begin{itemize}
  \item \emph{Beyond fail‑stop (gray/partial failures, correlation).} Introduce simple degradation/correlation knobs on edges and services; success = median $|\Delta\%|$ $\le$ 5\% (repl) and $\le$ 10\% (norepl) at mid‑range $p_{\mathit{fail}}$ across seeds; verify on targeted latency‑injection runs. \cite{huang_grey_failure}
  \item \emph{Source variants and fusion.} Demonstrate mesh‑only and IaC‑only discovery; enable optional fusion with per‑edge provenance; success = coverage~$\ge$~90\% of exercised edges and non‑worsening error vs.\ traces‑only. \cite{soldani2023offline}
  \item \emph{Generality.} Replicate the study on two additional applications and report cross‑app variance; success = overall Pearson $r\!\ge\!0.97$ and mid‑range bias within the thresholds above.
  \item \emph{Operationalization.} Integrate discovery into the observability stack to emit continuous “resilience posture” signals as topology changes; success = end‑to‑end runtime (discover\,$\to$\,estimate) $<$ 2 min and actionable alerts vetted with on‑call/SRE practice \cite{beyer_site_reliabili,li_enjoy_observe}, with alerts keyed to SLO targets when available.
  \item \emph{SLO-driven resilience triage.} Ingest SLO-as-Code files (when present) and treat them as a primary input for (i) automatically selecting/tagging critical entrypoints and (ii) comparing $\hat R_{\text{model}}(e,p_{\mathrm{fail}})$ against codified targets (availability/error budgets).
\end{itemize}

\noindent\textbf{Outlook.} The present results—high correlation and near‑zero mid\-‑range error under replication—indicate that automatically discovered, analyzable topology can provide accurate, low‑risk \emph{early} guidance and reduce the need for broad live chaos in advance, while reserving live experiments for the most critical or ambiguous scenarios. These planned extensions and external replications will elevate the work to a full paper by broadening validity, tightening error bounds, and demonstrating practical adoption in CI and observability.

\begin{acks}
The author thanks Alexander Zorkin from Tomsk State University for his crucial assistance in developing the CI/CD pipeline for the initial experiments and for his valuable feedback on an earlier version of this manuscript.
\end{acks}


\bibliographystyle{ACM-Reference-Format}
\bibliography{main}

\end{document}